\documentclass[a4paper]{IEEEtran}

\usepackage{ifpdf}
\usepackage[sort]{cite}
\ifCLASSINFOpdf
  \usepackage[pdftex]{graphicx}
  \usepackage{pdfpages}
  \graphicspath{{images/}}
\else
\fi

\usepackage{amsmath}
\usepackage{algorithmic}
\usepackage{array}
\usepackage{fixltx2e}
\usepackage{url}
\usepackage{amsmath,epsfig,amssymb,amsbsy,fixmath,mathrsfs,psfrag,epsf,enumerate,bm,color, epstopdf, pdfpages, bm}
\usepackage{graphicx}
\usepackage{amssymb}
\usepackage[caption=false]{subfig}
\usepackage{booktabs}
\usepackage{float}
\usepackage[section]{placeins}
\usepackage{comment}
\usepackage{textpos}

\begin{document}
\pagenumbering{gobble}	
	
\title{Full-Duplex Device-to-Device Collaboration\\ for Low-Latency Wireless Video Distribution}
\author{\IEEEauthorblockN{Mansour Naslcheraghi${{}^{1}}$, \textit{Member, IEEE} \\ Seyed Ali Ghorashi${{}^{1,2}}$, \textit{Senior Member, IEEE}, Mohammad Shikh-Bahaei${{}^{3}}$, \textit{Senior Member, IEEE}}\\
\IEEEauthorblockA{1. Department of Electrical Eng., Shahid Beheshti University, G.C. Tehran, Iran\\
2. Cyberspace Research Institute, Shahid Beheshti University, G.C. Tehran, Iran\\
3. Centre for Telecommunications Research, King's College London, UK\\
\url{m.naslcheraghi@mail.sbu.ac.ir}, \url{a_ghorashi@sbu.ac.ir}, \url{m.sbahaei@kcl.ac.uk}
}}
\maketitle
\begin{abstract}
 Growing demand for video services is the main driver for increasing traffic in wireless cellular data networks. Wireless video distribution schemes have recently been proposed to offload data via Device-to-Device (D2D) communications. These offloading schemes increase capacity and reduce end-to-end delay in cellular networks and help to serve the dramatically increasing demand for high quality video. In this paper, we propose a new scheme for video distribution over cellular networks by exploiting full-duplex (FD) D2D communication in two scenarios; scenario one: two nodes exchange their desired video files simultaneously with each other, and scenario two: each node can concurrently transmit to and receive from two different nodes. In the latter case, an intermediate transceiver can serve one or multiple users' file requests whilst capturing its desired file from another device in the vicinity. Analytic and simulation results are used to compare the proposed scheme with its half-duplex (HD) counterpart under the same transmitter establishment criteria to show the achievable gain of FD-D2D scheme in video content delivery, in terms of sum throughput and latency.
\end{abstract}
%
%
%\IEEEpeerreviewmaketitle 
%%{3cm}(4.1cm,-10.5cm)
%\begin{textblock*}{2cm}(3cm,-15.5cm)
%	\makebox[0.1\columnwidth]{$24^{\rm th}$ International Conference on Telecommunication (ICT 2017)}
%	
%\end{textblock*}


\IEEEpeerreviewmaketitle
\section{introduction}
Increasing demand for high data rate and live video streaming in cellular networks has attracted researchers' attention to cache-enabled cellular network architectures \cite{ref1}. These networks exploit D2D communications as a promising technology of 5G heterogeneous networks, for cellular video distribution. In a cellular content delivery network assisted by D2D communications and similarly in peer-assisted networks \cite{Nasreen}, user devices can capture their desired contents either via cellular infrastructure or via D2D links from other devices in their vicinity. Recently, several studies in both content placement policies and delivery strategies are conducted to minimize the downloading time, and to maximize the overall network throughput in terms of rate and area spectral efficiency. From the content placement perspective of view, contents can be placed on collaborative nodes formerly, either according to a predefined caching policy (reactive caching) \cite{ref2}, or more intelligently, according to statistics of user devices' interest (proactive caching) \cite{ref3}. %Network nodes in this system \cite{ref3} do not cache any content formerly; they can learn users' demand and thereby can predict users' future requests. 
The theoretical bounds for D2D caching network proposed in \cite{Theoretical2}, indicates that caching most popular contents in users' devices is optimal in almost all system regimes. % Furthermore, \cite{Tradeoffs} we have observed that the design of D2D architectures should balance the density of videos available through local caching with a proper use of time/frequency resources, so as to maximize the number of served requests. 
Cross-layer resource allocation methods are also investigated for supporting video over wireless in multiuser scenarios \cite{ref5}. It is shown that quality-aware resource allocation can improve video services in wireless networks. However, the conventional architectures of content delivery in both wireless cellular and D2D networks, are based on half-duplex (HD) transmission and to the best of our knowledge, full-duplex (FD) capability and its advantages have not yet been investigated in both wireless cellular video distribution and D2D caching systems. %FD radios can promise many advantages for future cellular networks.
Recent advances in FD radio design \cite{ref6}, materialized by advanced signal processing techniques that can suppress self-interference (SI) at the receiver, have enabled simultaneous transmission and reception over the same frequency band. From theoretical point of view, FD communication can potentially double the spectral efficiency of a point-to-point communication link, providing SI is entirely canceled. %Exploiting FD communication in a multi-node network can provide even higher throughput gains if multi-node interference is controlled through full-duplex MAC and network layer protocols. %Authors in \cite{ref9} prove that for slotted ALOHA MAC protocol, FD always outperforms HD in terms of probability of successful transmission. 
In this paper, we propose an FD-based scheme for D2D wireless video distribution. Details along with the main contributions are as follows: %The contributions in this paper are as follows:
 
\begin{itemize}
\item The proposed scheme has been investigated in two different scenarios: a) user devices operate in bidirectional FD mode in which two users can exchange data simultaneously at the same frequency and b) user devices can concurrently transmit to and receive data from two different nodes at the same frequency. i.e., an intermediate node can receive its desired content from one node and simultaneously serve for other user's demand at the same frequency.

\item We have analyzed throughput and delay in both scenarios and compared them against conventional HD systems.

\item In contrast with the works in the literature \cite{ref2}, where only one active node per cluster is considered, we consider D2D communication among multiple nodes in our proposed scheme. 
\item We have derived closed form expressions for FD/HD-D2D collaboration probabilities which previously obtained by numerical evaluations in \cite{MyIET}. 

\end{itemize}
The remainder of paper is structured as follows. In Section II system model is introduced. In section III, throughput analysis for the proposed FD-enabled cellular system is provided. In section IV simulation results are explained and conclusions are presented in section V.

\section{System model}
We consider a cellular network with a single cell, one base station (BS) and $n$ randomly distributed users (Fig.1 (a)) according to uniform distribution. Assuming that inter-cell interference is negligible or canceled out, analysis can be extended to multi-cell scenarios. We divide the whole cell area into logical equally sized square clusters (Fig. 1(a)) and neglect co-channel interference and neighboring cell users' influence, for the sake of simplicity. We consider an in-band overlay spectrum access strategy for D2D communications \cite{in-band}. Thus, there is no interference between cellular and D2D communications. All D2D communications are under full control of the BS. We also assume that SI cancellation allows the FD radios to transmit and receive simultaneously over the same frequency band. However, since all D2D pairs in all clusters share the same resource blocks, inter- and intra-cluster interference is taken into account. 
\begin{figure}[t]
	\centering
	\subfloat[Square cell with equal-sized clusters within the cell]{		\includegraphics[clip, width=0.2 \textwidth]{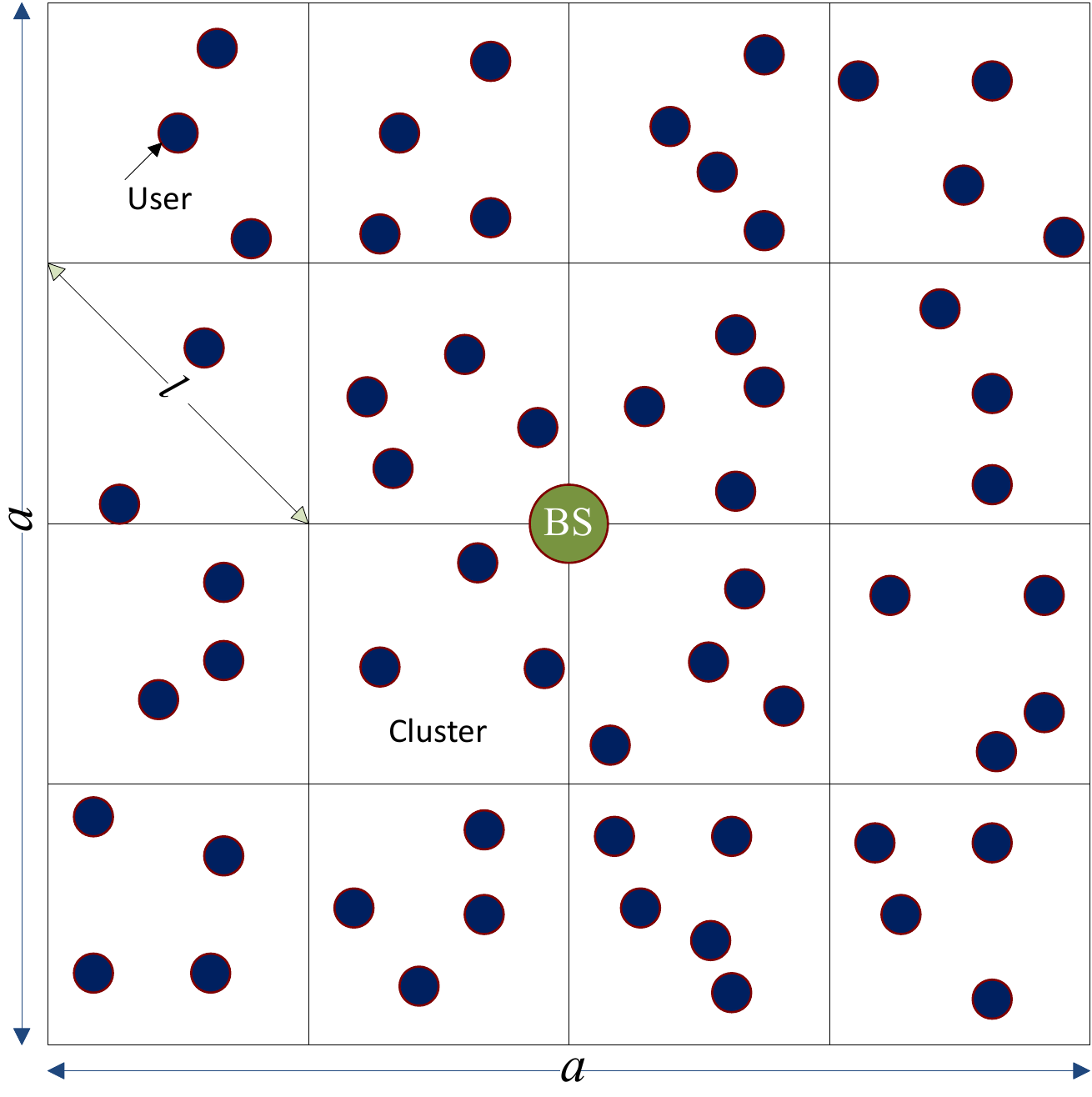}}
	\subfloat[Schematic of D2D communications]{
		\includegraphics[clip, width=0.23 \textwidth]{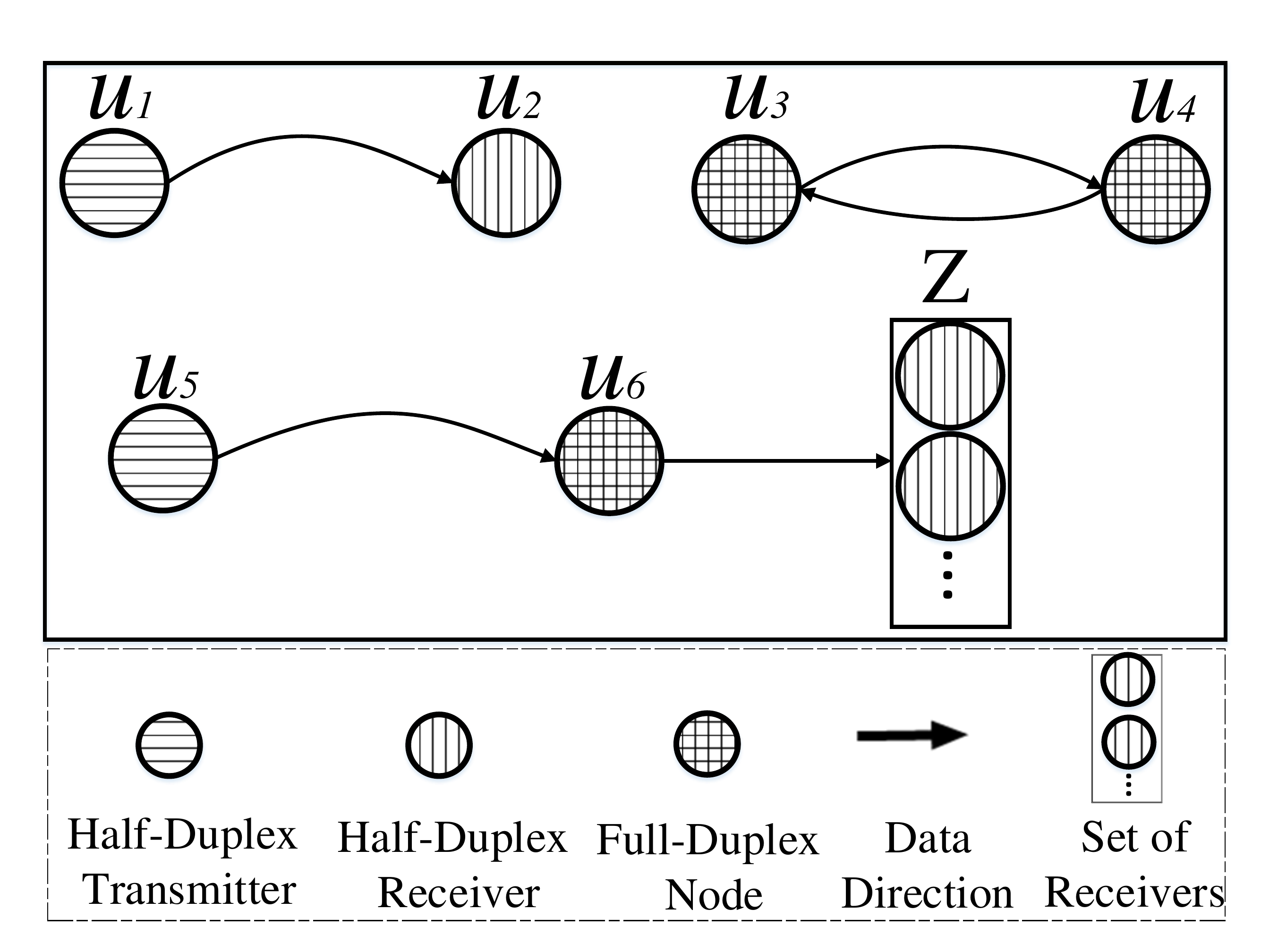}\quad\quad
	} \caption{System model and D2D communications graph}
\end{figure}

%\begin{figure}[t]
%	\centering
%	\begin{minipage}{.2\textwidth}
%		\centering
%		\includegraphics[width=1\linewidth, height=0.15\textheight]{SysModel-eps-converted-to.pdf}
%		\subcaption{Square cell with equal-sized clusters within the cell}
%	\end{minipage} \quad
%	\begin{minipage}{0.2\textwidth}
%		\centering
%		\includegraphics[width=1.1\linewidth, height=0.14\textheight]{NewRandomGraph.pdf}
%		\subcaption{Schematic of D2D communications}
%	\end{minipage}
%\caption{System model and D2D communications graph.}
%\end{figure}

Denote the set of popular video files as $\bm{V} = \{ {v_1},{v_2},...,{v_m}\}$ with size $m$. We use Zipf distribution for modeling the popularity of video files and thus, the popularity of the cached video file $v_s$ in user $u_{\omega}$, denoted by ${f_{\omega s}}$, is inversely proportional to its rank, i.e., $f_{\omega s}=\left( s^{\gamma_r}  \sum\nolimits_{g=1}^{m} g^{-\gamma_r} \right)^{-1}, \begin{array}{*{20}{c}}
{}&{1 \le s \le m}
\end{array}$. The Zipf exponent $\gamma _r$ characterizes the distribution by controlling the popularity of files for a given library size $m$. Contents are placed in users' caches in advance according to a caching policy in which each user with a considerable storage capacity can cache a subset of files $\bm{F_{\ell} \subset V}$ from the library, i.e., $\bm{F_{\ell}} = \{ {f_{\ell1}},{f_{\ell2}},...,{f_{\ell h}}\}$, $h \le m$. We assume that there is no overlap between users' caches, i.e., ${\bm{F}_p}\mathop  \cap \limits_{p \ne q} {\bm{F}_q} = \phi $. Each user randomly requests a video file from the library according to Zipf distribution. Technically, to schedule and establish a D2D connection, necessary signaling messages are needed to be exchanged between D2D pairs and the BS \cite{D2DDiscovery}. However, the signaling mechanisms do not affect our analysis in this work. Hence, we adopt the protocol model of \cite{ProtocolModel} to setup D2D communications, which is based on a distance threshold; A pair of users/devices $(u_i, u_j)$ can potentially initiate a D2D communication for video file transfer providing that the distance between $u_i$ and $u_j$ is less than a threshold ($l$ in Fig. 1(a)) and one of them finds its desired video file in the other device. %We expect that FD-enabled D2D collaborator nodes increase the number of active links inside a cluster and thereby increase the number of satisfied users whilst decreasing the latency in accessing video files. 
%Given a FD-enabled D2D transmitter, multiple links can be concurrently activated. 
Fig. 2(b) illustrates the schematic of typical D2D communication graphs inside a cluster. Each user generates a random request according to the Zipf distribution. BS is assumed to be aware of all contents in the users' caches. We define a directed edge from $u_i$ pointing to $u_j$ if the user $u_j$ requests a file that has been previously cached by $u_i$. Since we assume that each user can make only one request (as shown in Fig. 1(b)), there will be at most one incoming link to the user node and one or multiple outgoing links from the user node. In this system, no data is relayed over multiple hops, which means any transmission(s) from one node to another node(s) corresponds to delivering a different video content. It is also possible that some users demand for the same video content which is previously cached by one user. For instance the users in set $\bm{Z}$ demand for the same video content from user $u_6$ (Fig. 1(b)). The number of users in set $\bm{Z}$ depends on the popularity of the video content which is desired by these users. As can be seen in Fig. 1(b), there are two different possible configurations for FD collaboration; i) bi-directional full-duplex (BFD) mode, in which two users exchange their desired video content and ii) three node full-duplex (TNFD) mode, in which an intermediate node can receive its desired video content from one node and simultaneously serve for another user(s)' demand (see $u_6$ in Fig. 1(b)).

\section{Analysis}
Both analog and digital SI cancellation methods can be used to partially cancel the SI. However, in practice, it is difficult or even impossible to cancel the SI perfectly. We assume that all users transmit with power ${P_t}$. The SI in FD nodes is assumed to be canceled imperfectly with residual self-interference-to-power ratio $\beta$ and hence, the residual SI is $\beta P_t$. The parameter $\beta$ denotes the amount of SI cancellation, and $10{\log _{10}}\beta$ is the SI cancellation in dB. When  $\beta  = 0$, there is perfect SI cancellation, while for $\beta  = 1$, there is no SI cancellation. Thus, the signal-to-interference-plus-noise ratio (SINR) at receiver $u_j$ due to transmitted signal from $u_i$ can be written as 

\begin{equation}
\label{SINR formula}
\textup{SINR}_{i \to j} = \frac{{{P_t}{h_{ij}}d_{ij}^{ - \alpha }}}{{{\sigma ^2} + \sum\nolimits_{z \in \Phi \backslash \{ i\} } {{P_t}{h_{zj}}d_{zj}^{ - \alpha }}  + \chi \beta {P_t}}},
\end{equation}
where ${\sum\nolimits_{z \in \Phi \backslash \{ i\} } {{P_t}{h_{zj}}{d_{zj}}} }$ is total inter- and intra-cluster interference due to the nodes in set $\Phi$, which is the set of concurrent transmitting nodes. Backslash in eq. (\ref{SINR formula}) implies that the node $u_i$ is excluded from transmitters. ${{h_{ij}}}$ and ${{h_{zj}}}$ are the fading power coefficients with exponential distribution of mean one, corresponding to the channel between transmitter $u_i$ and receiver $u_j$, and interferer $u_z$, respectively. $d_{ij}$ denotes the Euclidean distance between transmitter $u_i$ and receiver $u_j$ inside the cluster. $\alpha$ is the path loss exponent. A white Gaussian noise with power ${{\sigma ^2}}$ is added to the received signal. $\chi$ denotes collaboration mode; $\chi=0$, when user $u_i$ operates in HD mode, and $\chi=1$, when it operates in FD mode.

\subsection{Collaboration Probability}
For given $k$ users which randomly fall inside a cluster and given $h$ number of cached contents for each user inside the random cluster $c$, we define popularity of cached contents within the cluster as ${{\rho}_{c}} = \sum\nolimits_{i = 1}^k {{\rho_{{u_i}}}}$, where $\rho_{u_i} = \sum\nolimits_{{f_{is}} \in {F_i }} {{f_{is}}} $ is the popularity of cached contents by user $u_i$. For the $i$th user, $u_i$, we define two parameters $P_{ai}$ and $P_{bi}$ as follows; $P_{ai}$: the probability that $u_i$ cannot find its desired content within cluster. $P_{bi}$: the probability that user $u_i$ can serve for other user(s)' request(s). Since all requests are identically distributed and independent (i.i.d) at each user, given $k$ users inside the cluster, the probability that $u_i$ operates in HD mode is
\begin{equation}
\label{PHD ui|k}
P_{{u_i}|k}^{\textup{HD}} = {P_{ai}}{P_{bi}}.
\end{equation}
Similarly, the probability that $u_i$ operates in FD mode is
\begin{equation}
\label{PFD ui|k}
P_{{u_i}|k}^{\textup{FD}} = {(1-P_{ai})}{P_{bi}}.
\end{equation}
However, the probability of making HD-D2D and FD-D2D connections depends on parameter $k$. The probability that $u_i$ can collaborate in HD or FD mode is
\begin{equation}
\label{P_ui^delta}
\mathcal{P}_{{u_i}}^{\delta} = \sum\limits_{k = 0}^n {P_{{u_i}|k}^{\delta}\Pr [K=k]}, 
\end{equation} 
where $\delta  \in \{ \textup{HD}, \textup{FD}\}$ is the operation mode, $\Pr [K=k]$ is the probability that there are $k$ users in the cluster. Since the distribution of users is assumed to be uniform within the cell area, the number of users  in the cluster is a binomial random variable with parameters $n$ and $\frac{{{l^2}}}{{2{a^2}}}$, i.e., $K = B(n,\frac{{{l^2}}}{{2{a^2}}})$, where ${\frac{{{l^2}}}{{2{a^2}}}}$ is the ratio of the cluster area to the cell area. Hence, the probability that $k$ users fall inside the cluster is
\begin{equation}
\label{Pr[K=k]}
\Pr [K = k] = \left( {\begin{array}{*{20}{c}}
	n\\
	k
	\end{array}} \right){\left( {\frac{{{l^2}}}{{2{a^2}}}} \right)^k}{\left( {1 - \frac{{{l^2}}}{{2{a^2}}}} \right)^{n - k}}.
\end{equation}
The probability that $u_i$ can find its desired file inside the cluster and cannot find on its own cache (i.e., we exclude self-request\footnote{self-request takes place when the user finds its desired file in its own cache.} from user $u_i$), can be written as
\begin{equation}
\label{Pai}
{P_{ai}} = {\rho_{c}} - {\rho_{{u_i}}}.
\end{equation}
We define $Q_{{u_i}}(x)$ which determines the probability that $u_i$ can serve $x$ number of users' requests inside the cluster. The number of users demanding for a content which is cached by $u_i$ is a binomial random variable with parameters $k-1$ and $\rho_{u_i}$, i.e.,
\begin{equation}
\label{Pserve(x)}
Q_{{u_i}}(x) = \left( {\begin{array}{*{20}{c}}
	{k - 1}\\
	x
	\end{array}} \right){\left( {{\rho_{{u_i}}}} \right)^x}{\left( {1 - {\rho_{{u_i}}}} \right)^{k - 1 - x}}. \begin{array}{*{20}{c}}
{}&{k \ge 2}
\end{array}
\end{equation} 
It is clear that for $k < 2$, $Q_{{u_i}}(x)=0$, which implies that there is no user's demand for cached content by $u_i$. And finally, $P_{bi}$ can be written as
\begin{equation}
\label{Pbi}
{P_{bi}} = \sum\limits_{x = 1}^{k - 1} {Q_{{u_i}}(x)}.
\end{equation}
By substituting eqs. (\ref{Pai}, \ref{Pbi}) in eqs. (\ref{PHD ui|k}, \ref{PFD ui|k}), and eqs. (\ref{PHD ui|k}, \ref{PFD ui|k}) in eq. (\ref{P_ui^delta}), respectively, we get the final mathematical expressions for HD and FD collaboration probabilities. 
\begin{align}
\label{HD final expression}
\mathcal{P}_{{u_i}}^{\textup{HD}} = \sum\limits_{k = 0}^n  {\left( {\left( {1 - \left( {{\rho_{c}} - {\rho _{{u_i}}}} \right)} \right)\sum\limits_{x = 1}^{k - 1} {Q_{{u_i}}(x)} } \right)} \Pr [K = k].
\end{align}
\begin{equation}
\label{FD final expression}
\mathcal{P}_{{u_i}}^{\textup{FD}} = \sum\limits_{k = 0}^n {\left( {\left( {{\rho_{c}} - {\rho_{{u_i}}}} \right)\sum\limits_{x = 1}^{k - 1} {Q_{{u_i}}(x)} } \right)} \Pr [K = k].
\end{equation}
Denoting $\mathcal{P}_{{u_i}}^{\textup{self}}$ as the probability that user $u_i$ finds its desired content on its own cache, By substituting eqs. (\ref{HD final expression}) and (\ref{FD final expression}) in $\mathcal{P}_{{u_i}}^{\textup{FD}} + \mathcal{P}_{{u_i}}^{\textup{HD}} + \mathcal{P}_{{u_i}}^{\textup{self}} = 1$, the probability that node $u_i$ demands for a file which is cached by itself is  
\begin{equation}
\mathcal{P}_{{u_i}}^{\textup{self}} = 1 - \sum\limits_{k = 0}^n {\left( {\sum\limits_{x = 1}^{k - 1} {{P_{{u_i}}^{serve}}} } \right)} \Pr [K = k].
\end{equation}

\subsection{Throughput Analysis}
We focus on a typical random cluster $c$ (representative cluster) and derive system sum throughput for this cluster. We obtain the ergodic capacity of the link associated with D2D pair ($u_i$,$u_j$), which is defined by ${C_{i \to j}} = WE[{\log _2}(1 + \textup{SINR}_{i \to j})]$, where, $W$ is the bandwidth for D2D link. For the wireless D2D network described in section II, the expected value of the throughput of the system due to establishing node $u_i$ in $\delta$ mode can be written as
\begin{equation}
T_{{u_i}}^{\delta} = \mathcal{P}_{{u_i}}^{\delta}C_{{u_i}}^{\delta},
\end{equation}
where $\mathcal{P}_{{u_i}}^{\delta}$ is the collaboration probabilities for $\delta$ mode, which is derived in equations (\ref{HD final expression}) and (\ref{FD final expression}). $C_{{u_i}}^{\delta}$ is achievable capacity by establishing node $u_i$ in $\delta$ mode and can be calculated as
\begin{equation}
\label{C HD ui}
C_{{u_i}}^{\textup{HD}} = \sum\limits_{{u_j} \in A} {WE[{{\log }_2}(1 + \textup{SINR}_{i \to j})]},
\end{equation}
\begin{align}
\label{C FD ui}
C_{{u_i}}^{\textup{FD}} =& WE[{\log _2}(1 + \textup{SINR}_{o \to i})] \notag\\&+ \sum\limits_{{u_{j}} \in B} {WE[{{\log }_2}(1 + \textup{SINR}_{i \to j})]}, 
\end{align}
where $A$ and $B$ are the set of users which are connected to $u_i$ in HD and FD modes respectively. First term in eq. (\ref{C FD ui}), i.e., $WE[{\log _2}(1 + \textup{SINR}_{o \to i})]$, determines the ergodic capacity for the link through which $u_i$ receives its desired file in FD mode from $u_o$ (TNFD mode). Showing the set of established nodes inside the random cluster $c$ is by $\bm{\Psi}  = \{ {u_1},{u_2},...,{u_{\tau}}\}$, the sum throughput of the respective cluster can be written as
\begin{equation}
\eta _c^\delta  = \sum\limits_{{u_i} \in \Psi } {T_{{u_i}}^\delta }.
\end{equation}

\subsection {Download Time}
As we described in section II, there are two full-duplex collaboration modes: TNFD and BFD. For better understanding the concept of download time in HD and FD modes, we use the D2D communication graphs shown in Fig. 1(b). 
\subsubsection{TNFD Mode}
consider $u_i$, $u_j$ and $\bm{Z} = \left\{ {{u_1},{u_2},...{u_k}} \right\}$ in which ${u_i} \notin \bm{Z}$, ${u_j} \notin \bm{Z}$.  
For a typical link between $u_i$ and $u_j$, and assuming that $u_i$ is transmitting video file $v_j$ to $u_j$, the experienced average download time $\theta_{i \to j}$ at $u_j$ can be defined as $\theta_{i \to j} = \frac{{{b_{v_j}}}}{{{C_{i \to j}}}}$, where $b_{v_j}$ is the number of bits for video file $v_j$ and $C_{i \to j}$ is the achievable ergodic capacity for transmitting link from $u_i$ to $u_j$. Similarly, for the set $\bm{Z}$, we have: ${\Theta} = \{\theta_{j \to 1}, \theta_{j \to 2},...,\theta_{j \to k}\}$ where, ${\theta_{j \to k}} = \frac{{{b_{{v_k}}}}}{{{C_{j \to k}}}}$. Due to random distribution of the users' locations, the ergodic capacity for all links associated with all users in set $\bm{Z}$ is not necessarily the same, hence ${\theta_{j \to p}} \ne {\theta_{j \to q}}$ for $p \ne q$. 
Since all users in set $\bm{Z}$ are demanding the same video content from $u_j$, the total average download time due to one transmission of user $u_j$ can be defined as 
\begin{equation}
\varpi  = \mathop {\max }\limits_{1 \le \lambda  \le k} ({\theta _{j \to \lambda }}),{\theta _{j \to \lambda }} \in \Theta. 
\end{equation}
Denoting $D_{u_j}^{HD}$ and $D_{u_j}^{FD}$ as the total experienced average download times by establishing $u_j$ in HD and FD modes, respectively, we have: 
\begin{equation}
D_{u_j}^{HD} = {\theta_{i \to j}} + \varpi, \quad D_{u_j}^{FD} = \max(\theta_{i \to j},\varpi).
\end{equation}
\subsubsection{BFD Mode}
In this mode, both users (i.e., $u_3$ and $u_4$ in Fig. 1(b)) exchange data simultaneously. Denoting ${\theta_{j \to i}}$ and ${\theta_{i \to j}}$ as the experienced download time for $u_i$ and $u_j$, respectively, the total average download time can be calculated as 
\begin{equation}
D^{HD} = {\theta_{j \to i}} + {\theta_{i \to j}}, \quad D^{FD} = \max(\theta_{j \to i},\theta_{i \to j}).
\end{equation}
In practice, the received and transmitted packets may have different lengths. Therefore, the transmission of all nodes will not end up at the same time. %The hidden node problem in FD transmissions 
Therefore, due to asymmetric data packets at the transmitter and receiver, this situation is referred to as ``the residual hidden node problem''. However, the node that finishes data transmission earlier can resolve this issue by transmitting busy tone signals until the other node completes its transmission \cite{HidenNodeProblem}. 

\section {Numerical Evaluations}
In this section, we provide Monte-Carlo simulation to evaluate the performance of our proposed FD-D2D caching system. We assume a single square cell as shown in Fig. 1(a). %Due to mobility, some users might exit the cell and the number of users within cell and consequently each cluster changes. Although the users can move, by applying a wraparound mechanism \cite{mobility}, the number of users in the cell is assumed to be fixed.
Simulation parameters are shown in Table 1. The proposed FD-scheme simulated based on the following scenarios:

\textit{Caching procedure}: %1000 files $\bm{V} = \{v_1, v_2... ,v_{1000}\}$ are considered in a library and 
Each user caches multiple files from the library, according to the described caching policy in section II. This procedure can be launched in the off-peak hours of the cellular network to avoid traffic load. 

\textit{Delivery procedure}: Users make and send their request to the BS randomly according to Zipf distribution and consequently the BS recognizes users' interests. Moreover, users' locations are known in advance for BS via channel state information (CSI) procedure. Hence, BS can predict potential D2D communications graphs (as such in Fig. 1(b)) for all clusters by having knowledge of users' caches, interests and locations. In each cluster, BS determines and establishes $\tau$ number of nodes associated with most popular cached contents. Since all D2D communications in all clusters use the same time-frequency resources, inter- and intra-cluster interferences are taken into account. Fig. 2 shows the probability that a node inside a cluster is in FD, HD or self-request mode. By increasing D2D collaboration threshold $l$, the expected number of users inside a cluster increases and, consequently, the expected number of nodes that collaborate in FD mode increases. %As we discussed in previous sections, some users can obtain their desired files through their own cache with zero delay (self-request users). 
As can be seen in Fig. 2, the probability that users can find their desired content on their own caches, decreases as $l$ increases, because for lower values of $l$, there are few users inside a cluster and these users previously stored high popular files. Hence, when the users inside a cluster make request according to Zipf distribution, there is a high probability that they make a request for a file that they have previously stored on their own caches. %In this situation, because of the low density of users inside a cluster, the probability is low that other users can make a request for high popular files. Therefore, the probability of making D2D communication is low, too. 
In contrast, as the density of users inside a cluster increases (i.e., higher values of $l$), the number of self-request users decreases. %This is because of high probability of making a request from one user to those most popular files to make a D2D communication. 
Fig. 3 shows the impact of the number of users ($n$) within the cell on the aforementioned probabilities. As can be seen from Fig. 3, the higher density of users within a cell, the higher is FD collaboration probability. Fig. 4 and Fig. 5, show the total average rate for FD-D2D and HD-D2D systems. Although the number of clusters increases at lower ranges of $l$ (we expect that the frequency reuse increases as well), nevertheless, the probability that clusters are of low density or that no D2D candidates are found therein, also increases. This can be interpreted as the fact that the probability of finding a user's desired file inside the cluster decreases when the node density decreases. As the number of clusters in the cell decreases, the frequency reuse decreases too. However, the probability that a user can find its desired file inside the cluster increases and hence the probability of making D2D communication increases. 
The impact of parameter $\tau$ ($\tau$ in Fig. 5) demonstrates that incorporating FD-enabled nodes, with multiple nodes establishments inside a cluster can improve the average gain in sum throughput by increasing the number of active D2D links. Alongside the considerable improvements in system sum throughput, the gain in frequency reuse in FD-enabled system is more accentuated for higher ranges of $l$. Fig. 6 illustrates the total average download time versus $l$. As we discussed in section II, there are three possible ways for the users to access their desired file; through conventional cellular infrastructure, via D2D collaboration, and by self-request. We define the download time as the delay incurred in downloading a file, i.e. the time between sending requests by the user till capturing the file. Download time for self-request case is zero and we exclude this case from calculations of download time. In the proposed FD-D2D system, each D2D receiver can download its desired file with zero waiting time. Fig. 6 shows the major impact of FD collaboration on decreasing the latency in downloading video files.
\begin{table}[b]
	\centering
	\caption {Simulation parameters} 
	\centering
	\begin{tabular}{|l|l|}
		\hline
		Parameter & Values\\
		\hline \hline
		$\bm{V}$ & Video Content Library\\
		$v_i$ & $i$th video content in library\\
		Number of users ($n$) & [10 1000]\\
		Cached contents per node ($h$) & 1, 3, 5\\
		Size of library ($m$) & 1000\\
		Zipf exponent ($\gamma_r$) &1, 1.6\\
		SI cancellation factor ($10{\log _{10}}\beta$) & -70 dB\\
		Number of established nodes ($\tau$) &  1, 2, 3\\
		D2D link bandwidth ($W$) & 1.2 MHz\\
		Background noise (${\sigma ^2}$) & -174 dBm/Hz\\
		Path loss exponent ($\alpha$) & 2.6\\
		Size of files & [5 50] MB\\
		User transmit power ($P_t$) & 23 dBm\\
		Cell size ($a$) & 1 km \\
		Log-normal shadow fading & 4 dB standard deviation\\
		Monte-Carlo iterations & 1000\\
		\hline
	\end{tabular}
	\label{notations}
\end{table}
 
\begin{figure}[t]{\vspace{+3mm}}
	\centering
	\includegraphics[width=0.37 \textwidth]{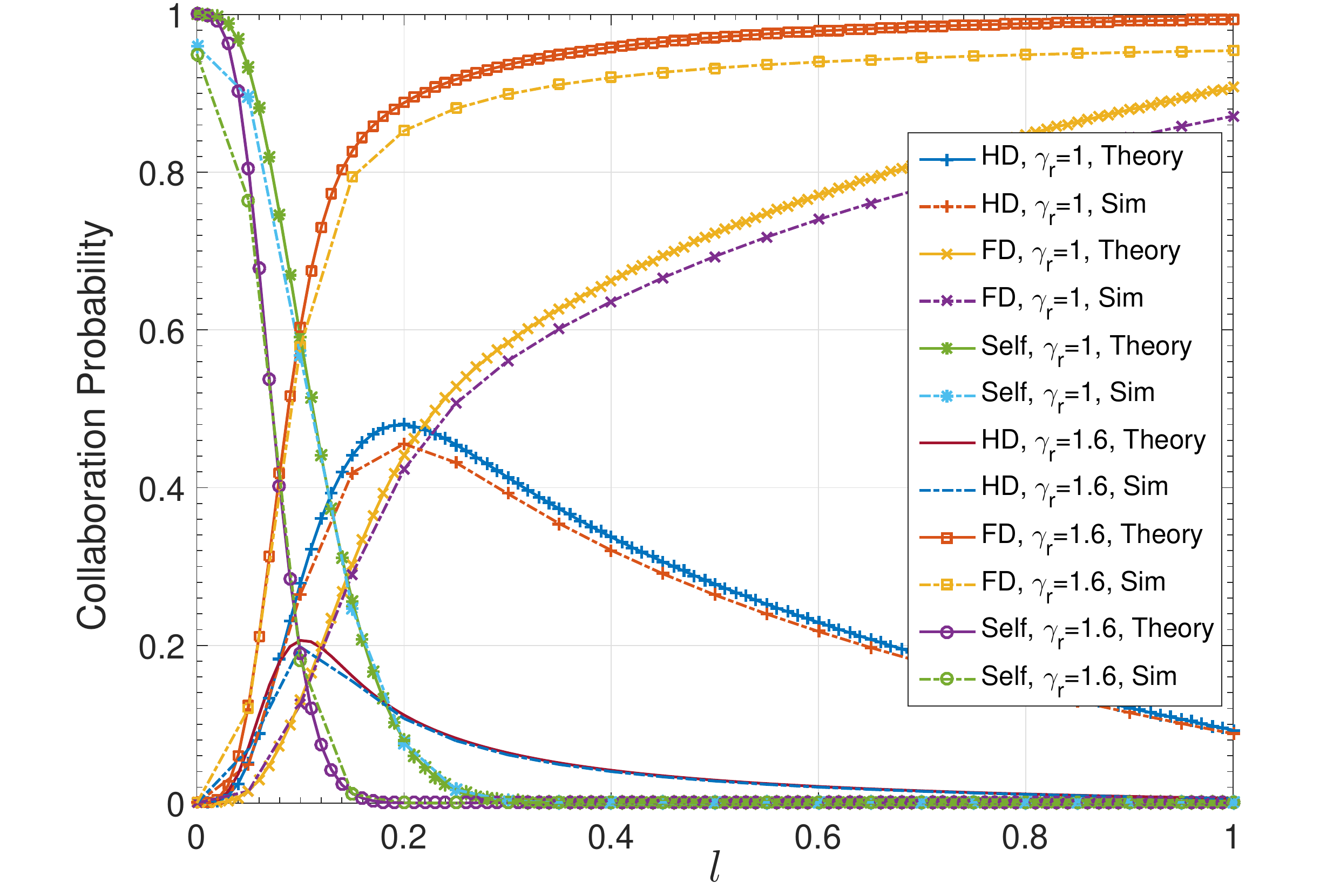}{\vspace{0mm}}
	\caption{Collaboration Probability versus $l$ for $n=500$ and $h=1$.}
\end{figure}
\begin{figure}[t]
	\centering
	\includegraphics[width=0.36 \textwidth]{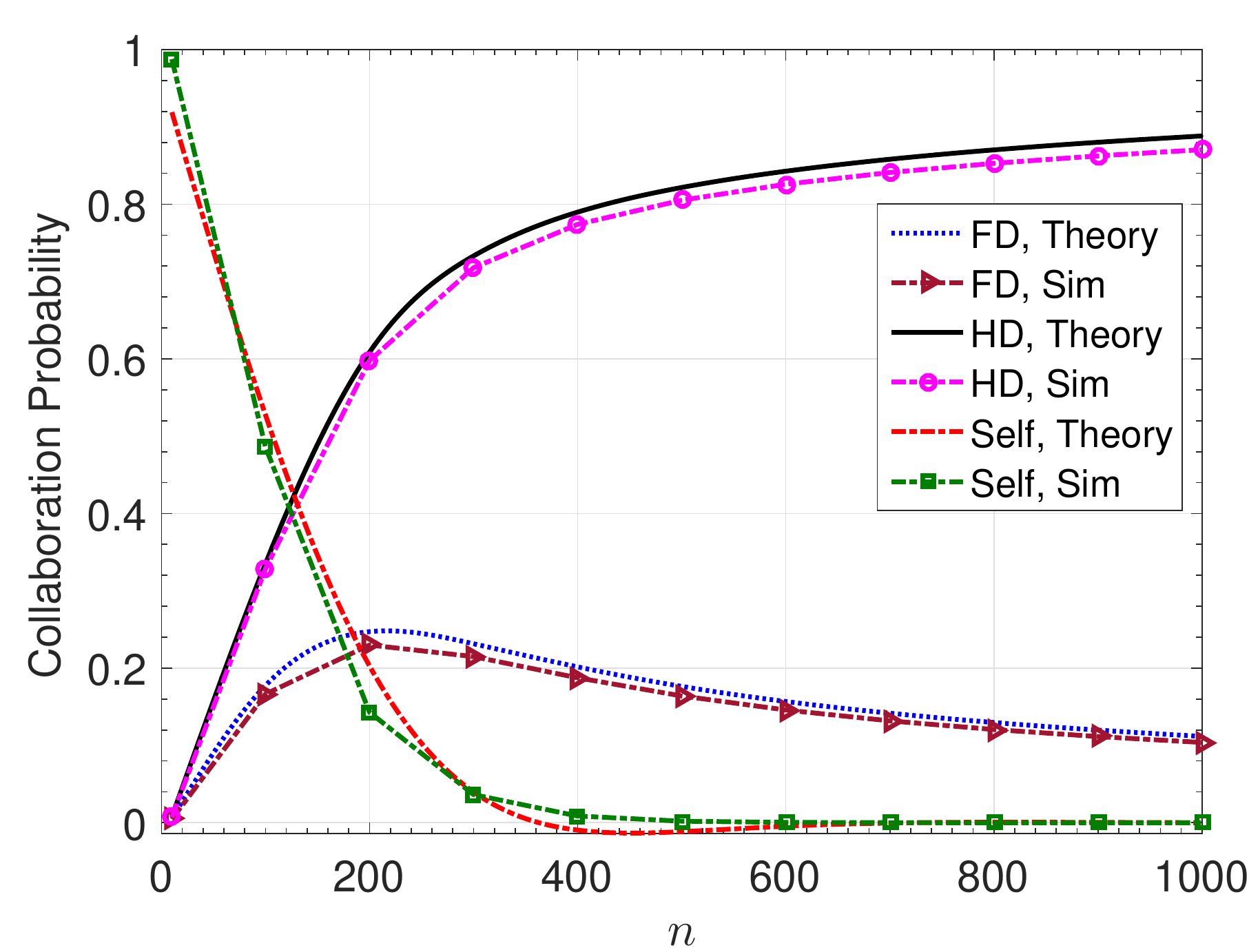}{\vspace{0mm}}
	\caption{Collaboration Probability versus $n$ for $\gamma_r=1.6$, $h=5$ and $l=0.2$.}
\end{figure}

\begin{figure}[!htb]
	\centering
	\includegraphics[width=0.38 \textwidth]{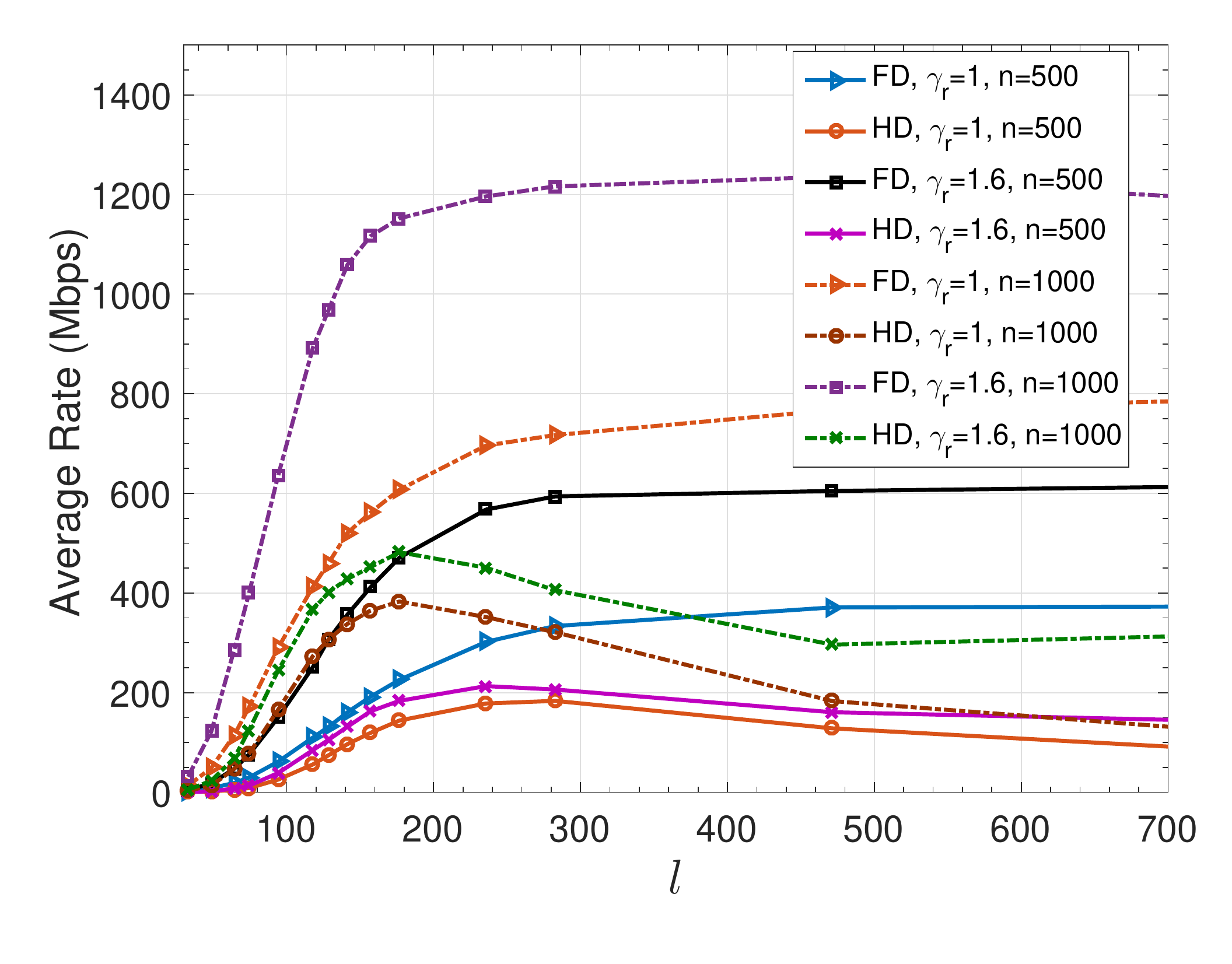}{\vspace{-3mm}}
	\caption{Average rate versus $l$ for $n=500$ and $h=1$ and $\tau=1$.}
\end{figure}

\section {CONCLUSION}
In this paper, we used full duplex radios on user devices to increase the throughput of video caching in cellular systems with D2D collaboration. We investigated FD-enabled networks by enabling FD radios only for D2D communications. Simulation results show that achievable throughput gain can increases in high intra- and inter-cluster interference conditions. We also showed that allowing full duplex collaboration can have a major effect on the quality of video content distribution by reducing download time compared to HD-only collaboration.

\begin{figure}[!t]
	\centering
	\includegraphics[width=0.36 \textwidth]{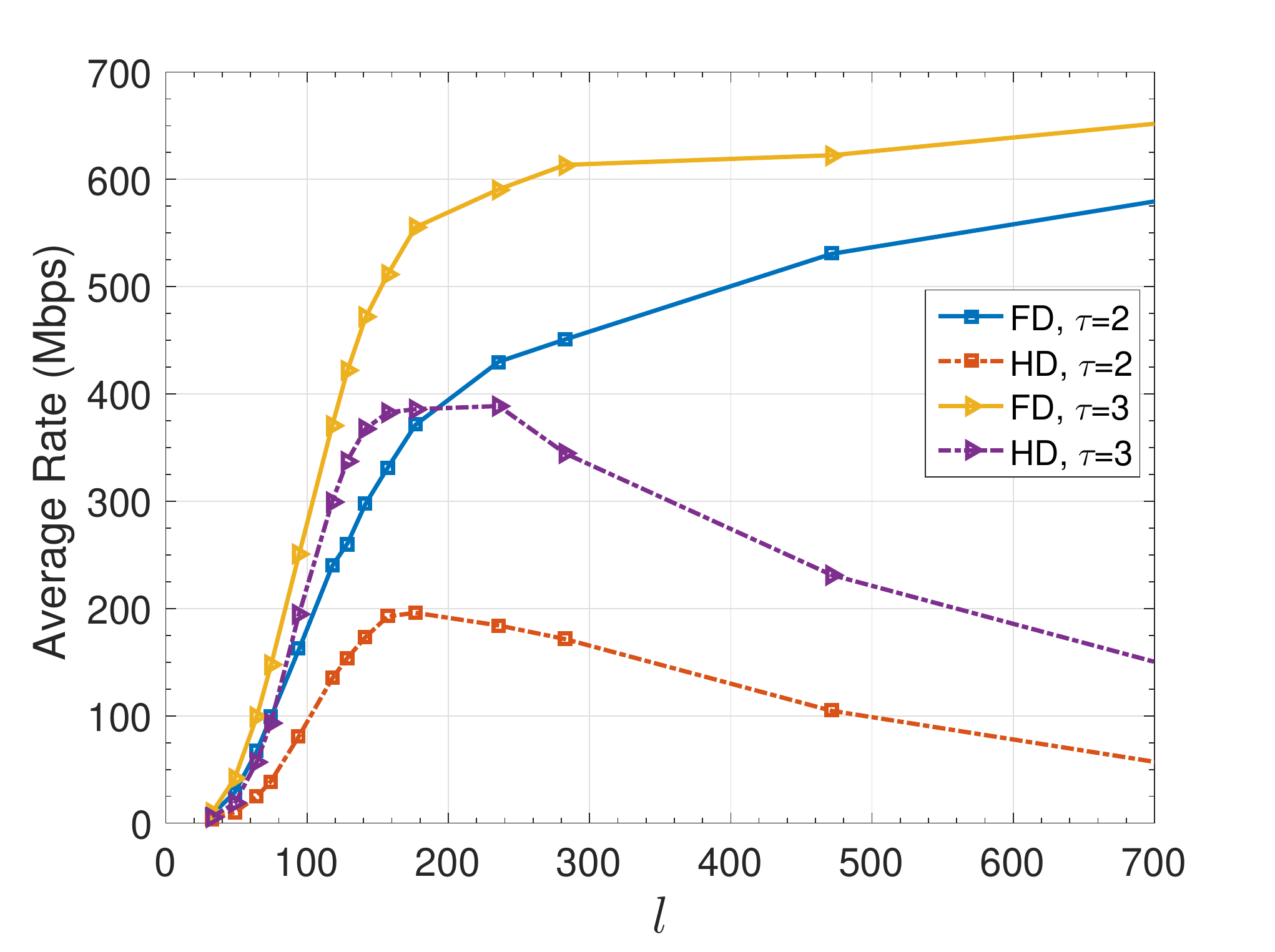}{\vspace{0mm}}
	\caption{Average rate versus $l$ for $h=3$, $\gamma_r=1$ and $n=1000$.}
\end{figure}

\begin{figure}[h]\vspace{0mm}
	\centering
	\includegraphics[width=0.38 \textwidth]{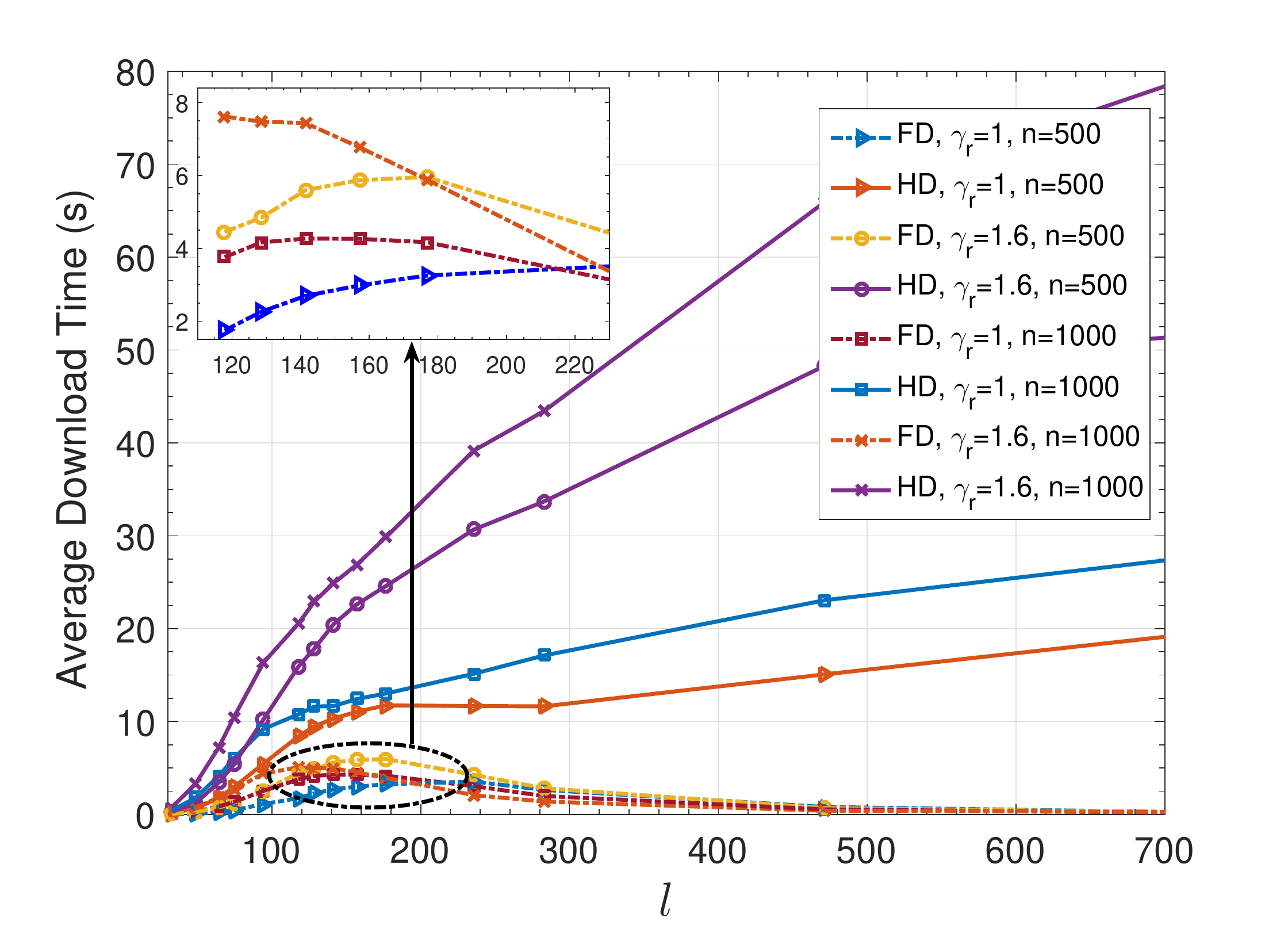}{\vspace{0mm}}
	\caption{Total average download time versus $l$ for $h=1$ and $\tau=1$.}
\end{figure}

\bibliographystyle{IEEEtran}

\end{document}